\documentclass{IEEEtran}
\IEEEoverridecommandlockouts
\usepackage{cite}
\usepackage{amsmath,amssymb,amsfonts}
\usepackage{algorithmic}
\usepackage{graphicx}
\usepackage{textcomp}
\usepackage{color}

\def\BibTeX{{\rm B\kern-.05em{\sc i\kern-.025em b}\kern-.08em
    T\kern-.1667em\lower.7ex\hbox{E}\kern-.125emX}}
\begin{document}
\title{A self-calibrating SI-traceable broadband Rydberg atom-based radio-frequency electric field probe and measurement instrument.}
\author{David Alexander Anderson, Rachel Elizabeth Sapiro, and Georg Raithel
%\thanks{This material is based upon work supported by the Defense Advanced Research Projects Agency (DARPA) and the Army Contracting Command- Aberdeen Proving Grounds (ACC-APG) under Contract Number W911NF-17-C-0007.}
\thanks{All authors are with Rydberg Technologies Inc., Ann Arbor, MI 48103 USA.  Corresponding author D. A. A. e-mail: dave@rydbergtechnologies.com.}}
%\IEEEmembership{Member, IEEE},
\maketitle

\begin{abstract}
We present a self-calibrating, SI-traceable broadband Rydberg atom-based radio-frequency (RF) electric (E) field probe (the Rydberg Field Probe or RFP) and measurement instrument (Rydberg Field Measurement System or RFMS).  The RFMS comprises an atomic RF field probe (RFP), connected by a ruggedized fiber-optic patch cord to a portable mainframe control unit with a computer software interface for probe RF measurement and analysis including real-time field and measurement uncertainty readout, and spectral RF waveform visualisation.  The instrument employs all-optical electromagnetically induced transparency (EIT) readout of spectral signatures from RF-sensitive Rydberg states of an atomic vapor for self-calibrated, broadband measurements of continuous, pulsed, and modulated RF fields.  The RFP exploits resonant and off-resonant Rydberg-field interactions to realize broadband RF E-field measurements at frequencies ranging from $\sim$10~MHz to sub-THz, over a wide electric-field dynamic range, with a single vapor-cell sensing element.  The RFMS incorporates a RF-field-free atomic reference as well as a laser-frequency tracking unit to ensure RFMS reliability and accuracy of the RF E-field measurement.  Atomic RF field measurement uncertainties reaching below 1\% are demonstrated.  We characterize the RFP and measure polar field patterns along primary axes of the RFP at 12.6~GHz RF, obtained by single-axis rotations of the RFP in the far-field of a standard gain horn antenna.  Field pattern measurements at 2.5~GHz are also presented.  The measured field patterns are in good agreement with finite-element simulations of the RFP. The data confirm that the atom-based RF E-field probe is well-suited for broadband isotropic RF measurement and reception. A calibration procedure and an uncertainty analysis are presented that account for deviations from perfectly isotropic response over $4\pi$ solid angle, which arise from asymmetric dielectric structures external to the active atomic measurement volume. The procedure includes contributions from both the fundamental atomic-spectroscopy measurement method and their associated analysis as well as uncertainty contributions due to material, geometry, and hardware design choices. The calibration procedure and uncertainty analysis yields a calibration (C) factor, used to establish absolute-standard SI-traceable calibration of the RFP.  Polarization pattern measurements are also performed, demonstrating RF-polarization detection capability with the instrument that can optionally be implemented simultaneously with E-field measurements.  RFP measurement capability for pulsed and modulated RF fields as well as direct, time-domain RF-pulse waveform imaging are demonstrated.  We conclude with a discussion of the practical use of the Rydberg atom-based RF E-field probe instrumentation in RF metrology towards the establishment of a new absolute (atomic) RF E-field measurement standard, application areas in RF measurement and engineering, and its value as a new quantum technology platform readily adaptable to specialized applications of Rydberg-based devices.

\end{abstract}


\section{Introduction}
\label{sec:introduction}
\IEEEPARstart{S}{ensors} and measurement devices for radio-frequency (RF) radiation at radio, microwave, sub-THz and THz frequencies enable capabilities essential to modern society with wide-ranging impact on industries spanning government and defense, telecommunications, electromagnetic compliance and safety, security, and medicine.  To date, RF field sensing and measurement has primarily relied on antenna technology to measure or receive RF electric (E) fields~\cite{Bassen.1983,Kanda.1987,Kanda.1993}.  Advances in antenna technology continue to provide improvements in RF capabilities.  Despite continuing advances, the very nature of traditional antenna technology, which is rooted in the driven oscillation of charges in a conductor induced by an incident RF electric field, imposes fundamental limits on the achievable accuracy, precision, and performance of probes and detectors for RF electric field measurement and sensing applications.

Atom-based quantum sensor technologies hold great promise for realizing capabilities beyond those achievable with traditional sensor technologies~\cite{MacFarlane.2003,Ludlow.2015,Battelier.2016,Alem.2017,Kitching.2018,Menoret.2018,Gerginov.2019}.  Recent advances in exploiting properties of individual atoms in highly-excited Rydberg states using optical electromagnetically induced transparency (EIT) in atomic vapors~\cite{Mohapatra.2007}, has afforded new capabilities in RF sensing, measurement, and imaging~\cite{Sedlacek.2012, Holloway2.2014, Anderson.2016, Anderson.2017}.  Rydberg atom-based RF electric field (E-field) sensing provides a combination of performance capabilities beyond what is possible with traditional antenna and other solid-state RF detectors.  This includes single-sensor ultra-broadband RF detection from HF to sub-THz~\cite{Holloway2.2014, Miller.2016,Wadearxiv.2016} and dynamic field ranges exceeding 120~dB, from field detection thresholds below 10~mV/m~\cite{Sedlacek.2012,Kuebler.2019} to high-intensity RF fields up to $\sim$10~kV/m, with atomic ionization limits at the MV/m level~\cite{Anderson.2017, Paradis.2019}.  Over a wide range of RF field amplitude and frequency, the Rydberg-based measurement method is rooted in physics models of the atom-field interaction that are dependent only on invariable atomic parameters and fundamental constants~\cite{Anderson.2014,Anderson.2016}.  This enables self-calibrated electric field measurements directly traceable to Planck's constant~\cite{Wood.2019} with atomic RF E-field measurement uncertainties reaching below 1\%~\cite{Miller.2016}, an improvement of nearly an order of magnitude over existing antenna standards~\cite{Hill.1990,Matoubi.1993}, holding promise to become a new global atomic RF measurement standard at National Metrology Institutes worldwide~\cite{Holloway2.2014, Humphreys.2017}.

In this work we present the first Rydberg RF E-field probe (Rydberg Field Probe or RFP) and measurement instrument (Rydberg Field Measurement System or RFMS) employing atom-based sensing using electromagnetically induced transparency (EIT) readout of spectral signatures from RF-sensitive Rydberg states in an atomic vapor~\cite{Anderson2.2017}.  The RFMS is a commercial instrument that comprises an atomic RF field probe (RFP), which houses a miniature atomic vapor-cell sensing element~\cite{RydbergTech} connected via a ruggedized fiber-optic patch cable to a portable rack-mounted control unit for remote probe operation and RF E-field measurement.  The RFMS is operated from a software user interface that provides real-time RF field measurement and uncertainty readout from the RFP, and RF-analysis features that include spectral and RF waveform visualization.  The RFMS measures RF fields by exploiting resonant and off-resonant Rydberg-RF field interactions in the RFP, together with RF-field-free atomic references and active laser-frequency tracking~\cite{Goncalves.2019} to ensure high reliability and accuracy in atomic RF E-field measurements.

This paper is organized into the following sections.  In Section~\ref{sec2} we provide a brief overview of Rydberg EIT readout in atomic vapors and RF E-field measurement.  In Section~\ref{sec3} we present and describe the RFP instrument and its operating principle, including the implementation of RF-field-free referencing and optical frequency tracking to achieve high reliability in precision RF E-field measurement and field determination methods for both linear and non-linear regimes of the atomic response with built-in compensation for perturbations of the RF field caused by the RFP-probe materials surrounding the atomic-vapor detection volume.  In Section~\ref{sec4} we characterize an RFP probe by performing polar field pattern measurements along three primary axes of the RFP at 12.6~GHz RF, obtained by single-axis rotations of the RFP in the far-field of a standard gain horn antenna, as well as field pattern measurements at 2.5~GHz RF.  The measured RFP field patterns provide atomic E-field measurement uncertainties below 1\%.  RF polarization detection and measurement with the RFP is also demonstrated.  In Section~\ref{sec4p5} Finite-element simulations of the RF field in the RFP are performed to quantify the effects of the RFP materials and design on the RF fields measured by the atoms, from which a calibration (C) factor is determined. With the C-factor, the RFMS provides absolute RF E-field measurements SI-traceable to Planck's constant and invariable atomic parameters.  In Section~\ref{sec5} we present an atomic RF field measurement uncertainty budget and analysis for the RFP relevant to SI-traceability of atomic RF probes and measurement tools in RF metrology.  In Section~\ref{sec6} we demonstrate RFP pulsed- and modulated-RF field measurement and direct time-domain RF-waveform detection and imaging.  In Section~\ref{sec7} we conclude with a discussion of the application of the RFP instrument in RF metrology and standards, RF engineering and measurement applications, and its use as a platform technology for other application-specific RF sensing, receiving, and measurement needs.

\section{Rydberg atom-based RF field sensing and measurement with EIT in atomic vapors}
\label{sec2}
\begin{figure}[!t]
\centerline{\includegraphics[width=\columnwidth]{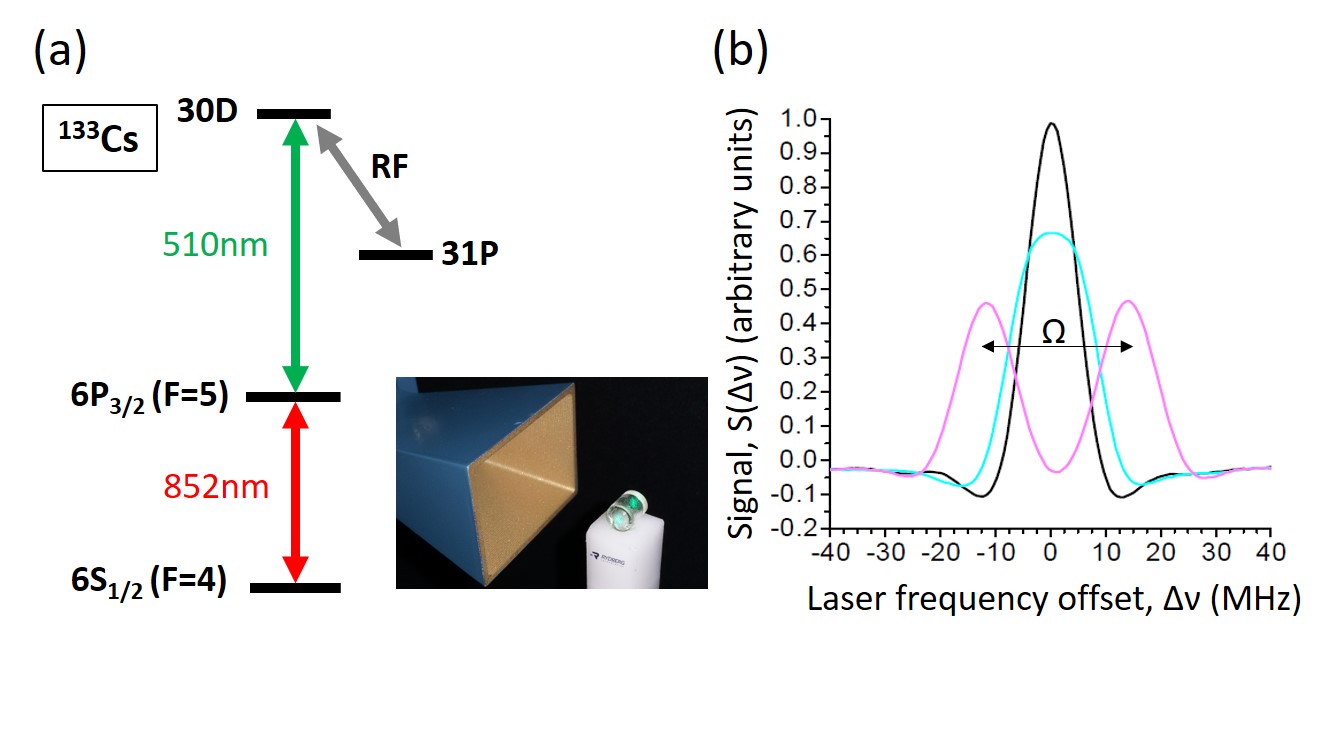}}
\caption{(a)Atomic energy-level diagram illustrating a two-photon Rydberg EIT optical readout scheme for a cesium vapor.  A miniature glass atomic vapor-cell sensing element in front of a standard horn antenna (inset). (b)  Optical readout from the atomic vapor of a Rydberg EIT resonance without RF (black curve) and in the presence of an RF field at a frequency near-resonant with an allowed transition between the optically excited Rydberg level and a second Rydberg level (magenta curve). The Autler-Townes (AT) splitting of the line yields the RF field strength.  In RF-fields too weak to AT-split the line, field-induced changes in the line shape can provide a measure for the RF field strength (cyan curve).}
\label{fig01}
\end{figure}

Rydberg atom-based sensing and measurement of RF fields utilizes optical electromagnetically induced transparency (EIT) readout of spectral changes from Rydberg states of an atomic vapor~\cite{Mohapatra.2007} that are sensitive to electric fields over a wide range of RF-field frequencies, amplitudes, and polarization~\cite{Sedlacek.2012,Holloway2.2014,Anderson.2016}.  Figure~\ref{fig01}(a) shows an atomic energy-level diagram illustrating a two-photon Rydberg EIT optical readout scheme for a cesium vapor.  The atomic (cesium) vapor is typically contained in a hermetically-sealed compartment with ports for optical access to the vapor; see, for example, the miniature glass vapor-cell sensing element in front of a standard horn antenna shown in the inset of the figure.  In the basic readout scheme, two optical laser fields couple atomic states to a high-lying Rydberg state (30D in Fig.~\ref{fig01}(a)), with a weak optical probe beam resonant with the first atomic transition between ground and an intermediate state, and a relatively stronger optical coupler beam tuned into resonance with a second atomic transition between the intermediate and Rydberg state.  When the coupler laser frequency is in resonance with the Rydberg state, an electromagnetically induced transparency (EIT) window opens for the probe beam through the vapor~\cite{Fleischhauer.2005,Berman}. Owing to the sensitivity of the atomic Rydberg levels to RF electric fields, the field-induced shifts and splittings of the Rydberg EIT signal enable an optical measurement for the RF field.  An example Rydberg EIT resonance is shown in Figure~\ref{fig01}(b) (black curve).  In the presence of a weak RF field at a frequency near-resonant with an allowed transition between the optically excited Rydberg level and a second Rydberg level of the atom, the EIT-detected atomic Rydberg line splits into a pair of Autler-Townes (AT) lines whose splitting is proportional to the RF electric-field amplitude (Figure~\ref{fig01}(c) (magneta curve)). In this linear AC Stark effect regime, the E-field is given by

\begin{equation}
E = \hbar \Omega /d,
\end{equation}

where $\Omega$ is the Rabi frequency of the RF-coupled atomic Rydberg transition (near-identical to the AT splitting measured optically in units $2\pi \times $Hz), $d$ is the electric dipole moment of the Rydberg transition in units Cm, and $\hbar=6.62606 \times 10^{-34}$~Js$/(2\pi)$ is Planck's constant.

From Equation~1 one can obtain an absolute, SI-traceable RF E-field measurement that is dependent only on invariable atomic parameters and fundamental constants.   Further, by changing the frequency of the coupler laser one can optically access different Rydberg levels that provide different RF field sensitivities and dynamic field ranges.  While the AT regime illustrated here provides an illustrative example of SI-traceable RF measurement with Rydberg EIT in vapors, the RFP implements a more generalized method that allows for measurements of RF fields at arbitrary frequencies over wide dynamic ranges, from low ($<$1~V/m) to high ($\sim$10~kV/m) RF fields.  The physics principles of this RF measurement method have been described in previous work~\cite{Anderson.2017,Anderson2.2017,Anderson.2018}.

\section{Rydberg RF Electric-Field Probe (RFP) and measurement system (RFMS)}
\label{sec3}

\begin{figure}[!t]
\centerline{\includegraphics[width=\columnwidth]{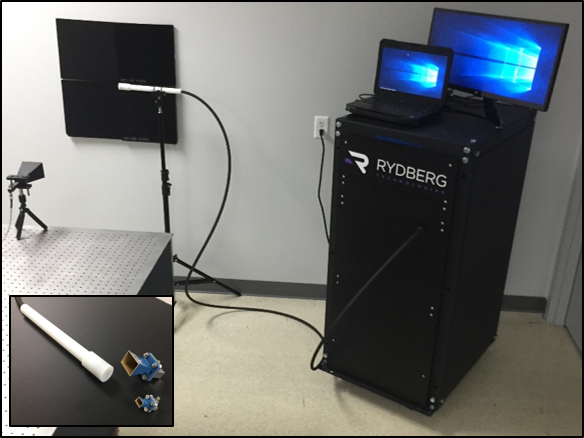}}
\caption{The Rydberg Field Probe (RFP) and mainframe control unit together composing the Rydberg Field Measurement system (RFMS). The RFP is shown in the inset next to two traditional horn antennas.}
\label{fig0}
\end{figure}

A picture of the RFMS, comprising an RFP and mainframe unit, is shown in Fig.~\ref{fig0}.  The RFP houses an atomic cesium vapor-cell sensing element that has a cylindrical geometry and 10-mm diameter and length.  The cell is unilaterally fiber-coupled, injecting 852-nm and 510-nm narrow-line laser beams overlapped and counter-propagating through the vapor, and returning the retro-reflected 852-nm light back to the instrument mainframe for optical readout of Rydberg resonances~\cite{Anderson.2019}.  The RFP vapor-cell sensing element is mounted on a probe rod and is connected to a portable mainframe by a ruggedized fiber-linked patch cable that is several meters long for remote operation. The mainframe contains all lasers and hardware that are automated via control software and a computer user-interface for RF field measurement with real-time RF field and uncertainty readout, RF signal analysis and visualization.  The RFP has a removable cap for protection of the sensing element during day-to-day operational use.  The RFP and its fiber-linked cable are fabricated out of hard dielectrics with small RF dielectric constants and loss-tangents to realize both a small footprint in RF field environments and mechanical robustness during operation.

\subsection{Operating principle for RF E-field measurements}

The RFP instrument realizes traceable measurements of RF E-fields by comparison of spectroscopic EIT signatures of RF-field-sensitive Rydberg states of atoms contained in the vapor-cell sensing element (see Section~\ref{sec2} and references) to absolute models of the Rydberg-atom response~\cite{Anderson.2016,Anderson2.2017}.  The patented method provides measurement capability of RF fields over a wide, continuous range of RF field frequencies, from $\sim$MHz to sub-THz, and of RF electric field amplitudes, extending from weak fields below 10~mV/m through a regime of moderate fields on the order of tens of V/m~\cite{Anderson.2014} to high-intensity RF fields above 10~kV/m~\cite{Anderson.2017}.  This field measurement method accounts for all non-linearities of the atomic response over the full RF range, which can be substantial for moderate-to-strong fields, thereby providing a self-calibrated linear E-field readout from the RFP over the full frequency and amplitude range of the RF radiation.  It is further noted that the method encompasses other limited approaches commonly implemented in laboratory experiments with Rydberg EIT RF field measurement.  These include the linear AC Stark effect, where an Autler-Townes (AT) splitting yields the RF electric field according to Eq.~(1). This approach is valid only for RF fields that are near-resonant with an RF-frequency-specific Rydberg-Rydberg transition, and the assumed linear relationship between field and AT splitting is accurate only over a limited dynamic field range, which can be as low as $\lesssim$10~dB~\cite{Sedlacek.2012, Holloway2.2014}. The RFMS also covers measurement approaches based of quadratic AC Stark shifts, which are suitable for measurements of continuously frequency-tunable RF fields that are off-resonant with any Rydberg-Rydberg transition. The AC shift approach is applicable to a dynamic range from $\sim 1$~V/m to $>$10~kV/m ($>$80~dB in intensity range)~\cite{Anderson.2018, Paradis.2019}, which is of considerable practical relevance, and which is wider than the dynamic range covered by the AT splitting approach. For simplicity and for proof-of-principle demonstrations, in this paper we perform the RFP RF field measurements, the RFP field pattern characterizations, the instrument's field determination method, and the measurement uncertainty analysis primarily in the AT and AC Stark shift regimes. In Refs.~\cite{Anderson.2016,Anderson2.2017, Sapiro.2019} we have described Rydberg atom-based RF sensing and measurement in the strong-field domain, which extends beyond the AT and AC-regimes.

\subsection{Frequency referencing and optical frequency tracking in the RFMS}

The accuracy of the RF E-field measurement with Rydberg EIT spectroscopy relies on the accuracy with which the optical frequency between the RF-altered atomic spectral features can be measured.  This is dependent on a laser-scan frequency calibration that is not a standard feature in available laser devices.  Further, the nature of the RF-induced spectral features also depends in part on the choice of atomic Rydberg state used for a given RF field measurement; for example, Rydberg S-states and D-states result in different spectral responses for similar RF field frequencies or amplitudes due to their magnetic substructure and other differences.  As a general solution, the RFMS employs an RF-field-free atomic reference and a scanning laser-frequency tracker in real-time during RFP operation that provide maximal versatility and ensure high reliability and accuracy in RF E-field measurement.

\begin{figure}[!t]
\centerline{\includegraphics[width=\columnwidth]{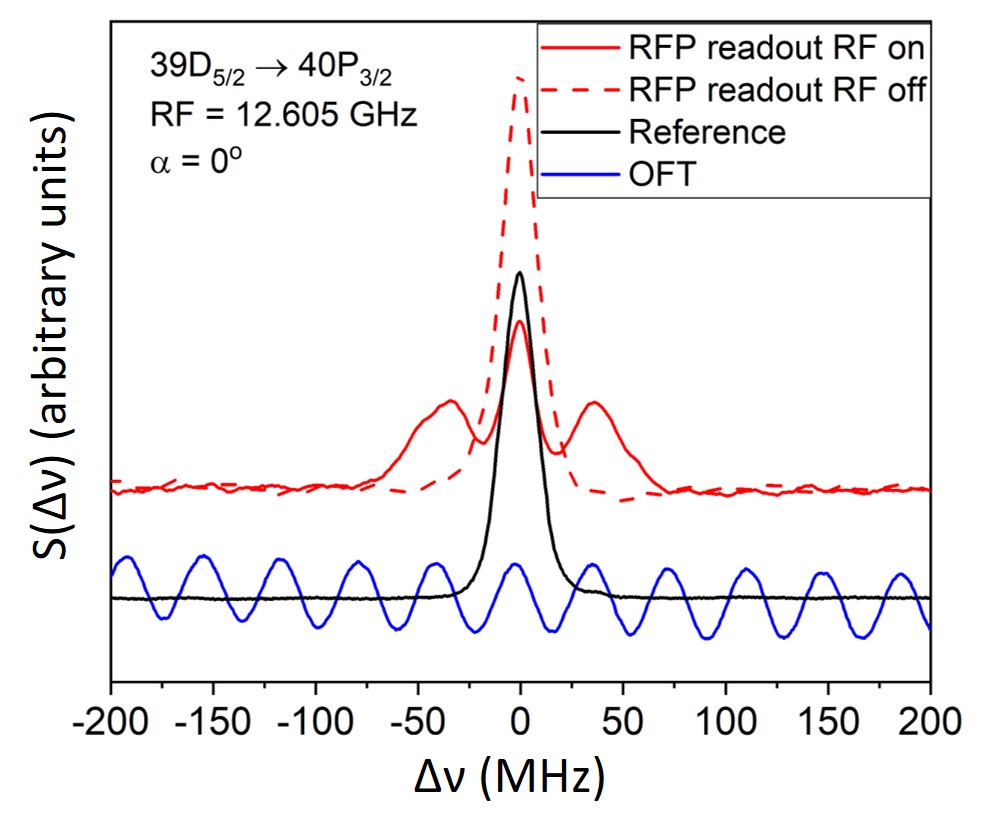}}
\caption{Spectroscopic and optical signals simultaneously collected by the RFMS during a 12.6~GHz RF E-field measurement with the RFP in the far-field of horn antenna emitter.  Shown are the EIT readout from the RFP with the RF field on (red, solid) and off (red, dashed), the EIT readout from an RF-field-free Rydberg reference unit housed within the RFMS mainframe (black), and the readout from an optical frequency tracker (OFT) in the RFMS mainframe (blue). The OFT signal provides a ``frequency-ruler'' with a calibrated period of (38.30 +/- 0.02)~MHz.}
\label{fig1}
\end{figure}

Figure~\ref{fig1} shows three example optical readout signals collected simultaneously by the RFP instrument during an RF E-field measurement. In the measurement, the coupler laser (510-nm laser; see Fig.~1) is scanned, and the recorded signals are displayed as a function of laser detuning. The signals are: (1) the spectroscopic Rydberg EIT readout from the RFP in the RF field of interest, (2) an RF-field-free atomic reference spectrum, and (3) periodic optical frequency markers for the laser scan derived from a calibration-free Optical Frequency Tracker (OFT; Rydberg Technologies Model OFT-NIR-19)~\cite{Goncalves.2019}.  The latter two signals are generated by internal devices in the instrument mainframe and ensure high reliability and spectroscopic accuracy for atomic RF E-field measurements, reaching absolute RF E-field measurement uncertainties at the 1\% level and below.  For illustration purposes, in Fig.~\ref{fig1} the RF field frequency and amplitude measured by the RFP are selected such that the atomic response is a resonant AT splitting whose value is approximately linear in the RF electric field, allowing for a simple and accurate determination of the RF E-field with low measurement uncertainty.  The OFT in the RFMS mainframe provides a frequency ruler for the scanned 510-nm laser with a calibrated fringe spacing (38.30$\pm$0.02~MHz in this work).  The OFT signal tracks the laser frequency in real-time during the coupler (510-nm) laser scans and provides an absolute, high-precision laser-frequency axis calibration. A calibrated laser-frequency axis is an important ingredient that allows the RFMS operating software to process the Rydberg EIT spectra measured within the RF field, and to report an RF field-amplitude reading to the RFMS user.

\subsection{RF field determination methods}

The RFP instrument measures RF E-fields by converting the optical readout of the field-modified atomic response from the vapor-cell probe to an E-field value. Spectroscopic features (observed as probe-beam transmission changes) are matched to pre-calculated features of the atomic response linked to invariable atomic properties and fundamental constants; as these features are unique, they can be corresponded to an E-field value via this comparison.  To perform SI-traceable, self-calibrated broadband RF measurements over a wide dynamic range, the RFP implements RF E-field determination methods across all atom-field interaction regimes with Floquet that includes AT and AC Stark shifts~\cite{Anderson2.2017}.  The RFMS implements dedicated spectral analyses for field-determination that account for both the varying atomic response in the different atom-field interaction regimes as well as broadening and alteration of the atomic spectral features in the readout due to RF field inhomogeneities that may be present in the atomic detection volume as a result of perturbations of the field by RFP materials.

\begin{figure}[!t]
\centerline{\includegraphics[width=\columnwidth]{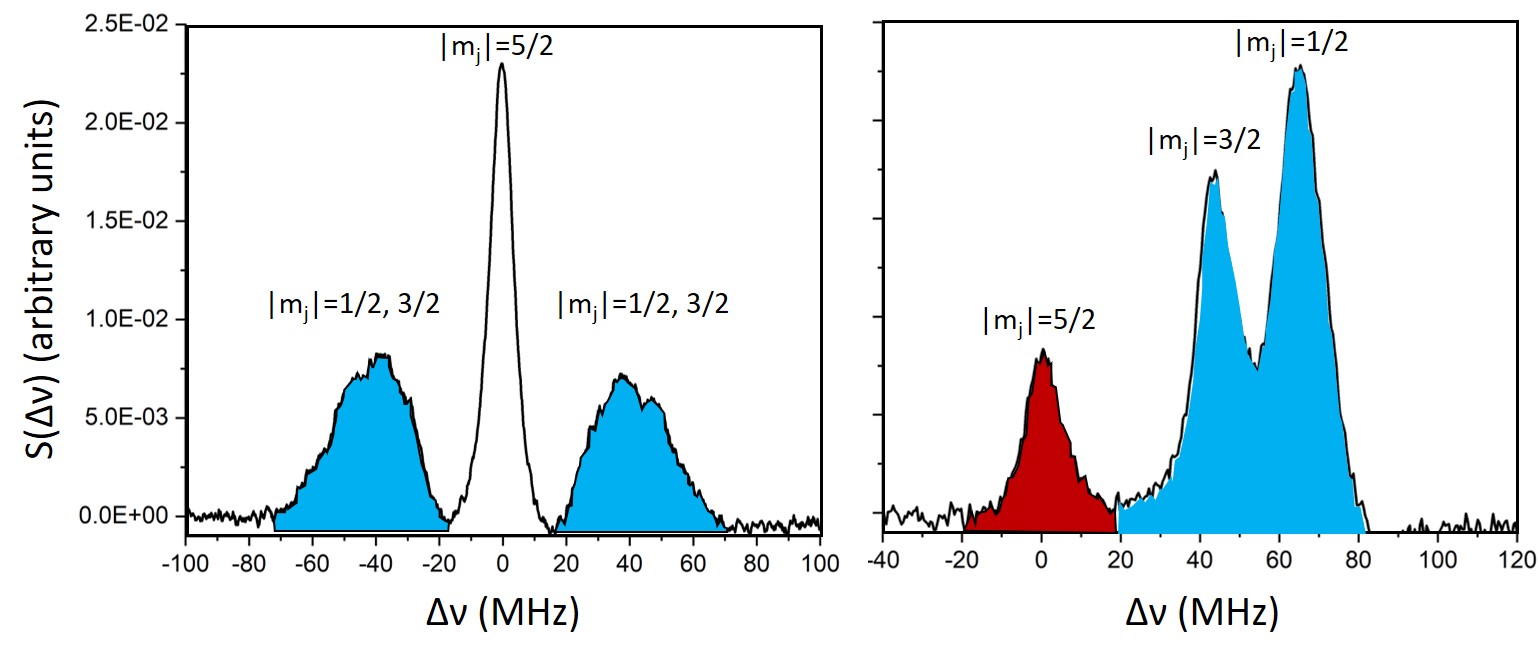}}
\caption{Left: Rydberg Field Probe (RFP) optical atomic spectrum $S(\Delta\nu)$ versus laser-frequency detuning $\Delta\nu$ showing Autler-Townes (AT) splitting for a measurement of a 12.6~GHz RF electric field resonant with the Cs $39D_{5/2}\rightarrow 40P_{3/2}$ transition. The regions over which the AT-split lines are integrated in order to obtain $\Delta \nu$ (see text) are shaded in blue. Right: off-resonant AC Stark shifts of the Cs $48D_{5/2}$ state for the measurement of a 2.5~GHz RF electric field. The two blue-shifted AC-Stark shifted lines belong to the magnetic sub-states $m_j=1/2$ and $3/2$ lines and are shaded in blue, while the $m_j=5/2$ line is shaded in red. }
\label{figATAC}
\end{figure}

Figure~\ref{figATAC} shows example RFP spectrum readouts for RF E-field measurements performed in the AT and AC-Stark regimes for a 12.6~GHz and 2.5~GHz RF E-field, respectively.  The methods for determining the E-field in the AT and AC-Stark regimes are similar. In both cases, a routine auto-locates the peaks in the EIT signals and employs a signal averaging approach to account for line-shape substructure.  Here, the spectral EIT signal $S(\Delta\nu)$ is integrated over the field-altered peak. From this integral, the average frequency shift, $\langle \Delta\nu \rangle $, of the RF-altered peaks is determined with respect to the field-free EIT spectrum,
\[ \langle \Delta\nu \rangle = \frac{\int \Delta\nu \, S(\Delta\nu) d\Delta\nu}{\int S(\Delta\nu) d\Delta\nu} \quad.\]
%The integration ranges, indicated by red and blue areas in Fig.~\ref{figATAC}, are determined by locating intermediate minima of the signal $S(\Delta\nu)$ between adjacent peaks and checking for by convergence of $S(\Delta\nu)$ with the noise floor.
In the AT regime, the average RF field in the RFP can be obtained from either side peak using the equation $\langle E\rangle =2h\langle \Delta \nu \rangle /(d_{\rm{rad}}d_{\rm{ang}})$, where $h$ is Planck's constant, $d_{\rm{rad}}$ is the radial matrix element of the Rydberg transition, and $d_{\rm{ang}}$ is the angular matrix element.  There, $E$ is the amplitude of the RF electric field amplitude, and its average $\langle E \rangle$ is over position within the RFP field-probe volume and over magnetic sub-states of the atoms, as appropriate.
The representative measurement example in Fig.~\ref{figATAC} (left plot) is in a low-E-field regime, where the AT-split lines contain known substructure of spectral components belonging to different values of the magnetic quantum number $m_j$ that do not separate.  As the angular matrix elements, $d_{ang}$, for $|m_j|=1/2$ and $|m_j|=3/2$ differ by 20\%, choosing one component for the E-field determination and ignoring the other would give a wrong result.  To account for this the RFMS approximates the angular matrix element by using an average over the relevant cases of $m_j$ ($|m_j|=1/2$ and $|m_j|=3/2$ for the case in Fig.~\ref{figATAC}) (left plot).

In the AC-Stark regime, $\langle E^2\rangle =4\times\langle \Delta \nu \rangle /\alpha_j$, where $\alpha_j$ is the AC polarizability, which depends on the Rydberg level, the $|m_j|$ state, and on the RF frequency. %The known dependence of $\alpha_j$ on frequency resides in the RFMS mainframe memory.
The AC-Stark-induced line shift provides a measurement of, equivalently, $\langle E^2\rangle$, the RMS value of the RF electric-field amplitude averaged over the RFP field-probe volume, and the intensity of the RF radiation. The RFMS field determination method in the AC-Stark regime is implemented similarly to the AT regime. An auto-location and peak integration routine applied to the AC-shifted and AC-split spectra again accounts for line-shape substructure and line overlaps that occur over the $>80$~dB-wide $\sim$1~V/m to $>10$~kV/m dynamic field range.  In the illustrative example shown in Fig.~\ref{figATAC} (right plot), the field determination routine is applied to AC-shifted and AC-split spectra of $nD_{5/2}$ Rydberg states of cesium that have $|m_j|= 1/2$ and $3/2$ lines slightly overlapped and together separated from the $|m_j|=5/2$ line.  Here, the field determination routine takes advantage of the higher field sensitivity of the $|m_j|=1/2, 3/2$ lines, which have large $\alpha$ and exhibit larger line shifts at low fields, compared to that of the $|m_j|=5/2$ line, which instead exhibits a stronger EIT signal, line changes and shifts at higher fields, for RF field measurement over the wide dynamic range afforded by the AC Stark regime.

The RFMS peak-integration approach described in this section is generalizable to any other field-detection method that involves comparing observed and calculated spectral features, including the most general case of Floquet calculations.  Throughout this discussion we have found the average peak position, $\langle \Delta \nu \rangle $, and used it to determine the average field $\langle E\rangle $ or the average of $\langle E^2\rangle$ (which is proportional to average RF intensity). Another possible field-determination approach would be to locate the $\Delta \nu$ values at which the signal $S(\Delta \nu)$ peaks, and to use those $\Delta \nu$ values to compute the RF electric field.  In this method, the $\Delta \nu$ values that correspond to the peak positions would be obtained by locating the actual maxima of the signal $S(\Delta \nu)$, or by performing local fits over peak regions in the signal and locating the maxima of these local fits. These most-common peak positions can be misleading, however: if there are field inhomogeneities internal to the RFP due to the RFP's materials and geometry, the most-common field may be a node or anti-node of an internal standing wave rather than a true measure of the RF field incident on the RFP. Such inhomogeneities can arise even in a measurement device much smaller than the RF wavelength due to dielectric boundary conditions. The average field across the entire detection region represents the incident field more faithfully. We find that the average field and the most-common field can differ by as much as 15-20\%. Additionally, even when no field inhomogeneities are present, the presence of multiple transitions with different RF-field-shifted values of $\Delta \nu$ due to magnetic substructure of the atoms can distort the shape of the spectroscopic peaks in $S(\Delta\nu)$. At low RF field levels, separate $|m_j|$ peaks cannot be resolved, but the shape of the unresolved compound peak depends on the detailed widths and strengths of its unresolved sub-components, which may further depend on RF polarization. Such effects would artificially skew a peak-finding or fitting method toward one of the unresolved sub-components contributing to the compound peak. Finding the average peak position using the above explained integration method, however, and then determining the E-field based on weighting contributions from the different $|m_j|$ components largely ameliorates this skew. Additionally, the averaging method is robust against RF standing-wave effects and inhomogeneities within the cell. %A detailed discussion of the field-determination method is presented in~\cite{Sapiro.2019}.

\section{RFP field pattern measurements}
\label{sec4}

The utility of an RF probe or sensor in measurement and receiving applications relies on calibration and validation of the directional dependence of the detector sensitivity to incident RF waves.  In this regard, the RFP atom-based RF sensing exhibits fundamental differences and performance advantages compared to antennas.  First, the RFP atom-based RF sensing method provides a rare case of a true isotropic receiver.  Unlike antennas, which due to electromagnetic boundary conditions cannot be built to radiate or receive in all directions, an atom is sensitive to an RF field incident from any direction.  This is due to the fact that the quantum structure of the atomic states is always altered in the presence of an incident RF field, and the RF-sensitive states can be accessed with an optical (or electronic) readout.  Second, the atom-based method optionally provides the capability of RF polarization detection simultaneously with, but independent of, electric field measurement.  This is due to the fact that shifts or splittings of the field-modified spectroscopic lines depend on the amplitude of the RF field, while the relative strengths of the spectroscopic lines depend on the orientation the incident RF field polarization relative to that of the optical polarization.  Independent (and simultaneous) RF polarization and field detection is not possible with an antenna, whose sensitivity to the field is intrinsically linked to the orientation of the RF field polarization relative to the conductive antenna structure.

We characterize the directional dependence of the RFP atom-based RF field probe by performing field and polarization pattern measurements of the RFP. To achieve this, we employ the methods described in Sec.~\ref{sec3} to determine RF electric fields and their uncertainties for a range of conditions, using AT splittings (Sec.~\ref{sec4} and~\ref{sec5}) and AC shifts (Secs.~\ref{sec5} and~\ref{sec6}).

\begin{figure}[!t]
\centerline{\includegraphics[width=\columnwidth]{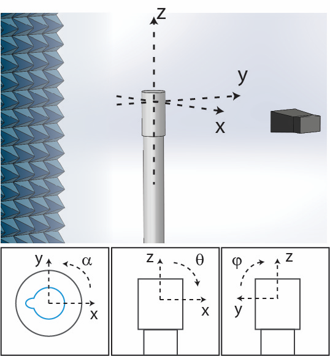}}
\caption{Schematic of the RFP field pattern measurement setup (top) and the RFP single-axis rotation angles $\alpha$, $\theta$, and $\phi$ about the primary axes $\hat{\bf{z}}$, $-\hat{\bf{y}}$ and $\hat{\bf{x}}$    (bottom).}
\label{fig2}
\end{figure}

Figure~\ref{fig2} shows an illustration of the measurement setup.  The RFP is placed at an initial position $(X,Y,Z)=(0,0,0)$~mm, with an uncertainty of $\pm 1$~mm in each component, defined to be the center of the 10~mm cylindrical vapor cell and optical detection volume. The orientation shown in Fig.~\ref{fig2} corresponds to $(\alpha, \theta, \phi)=(0,0,0)$. Under this condition, the vapor-cell's cylinder axis points along $\hat{\bf{z}}$, its stem along $-\hat{\bf{x}}$, and the linear optical polarizations inside the probe cell along $\hat{\bf{y}}$.  The RF source is placed at position $(X,Y,Z)=(325,0,0)$~mm, with an uncertainty of $\pm 5$~mm in each component. At the location of the probe, the RF field has a propagation vector pointing along $-\hat{\bf{x}}$ and a linear polarization pointing along $\hat{\bf{z}}$ .

\begin{figure*}[!t]
\centerline{\includegraphics[width=18cm]{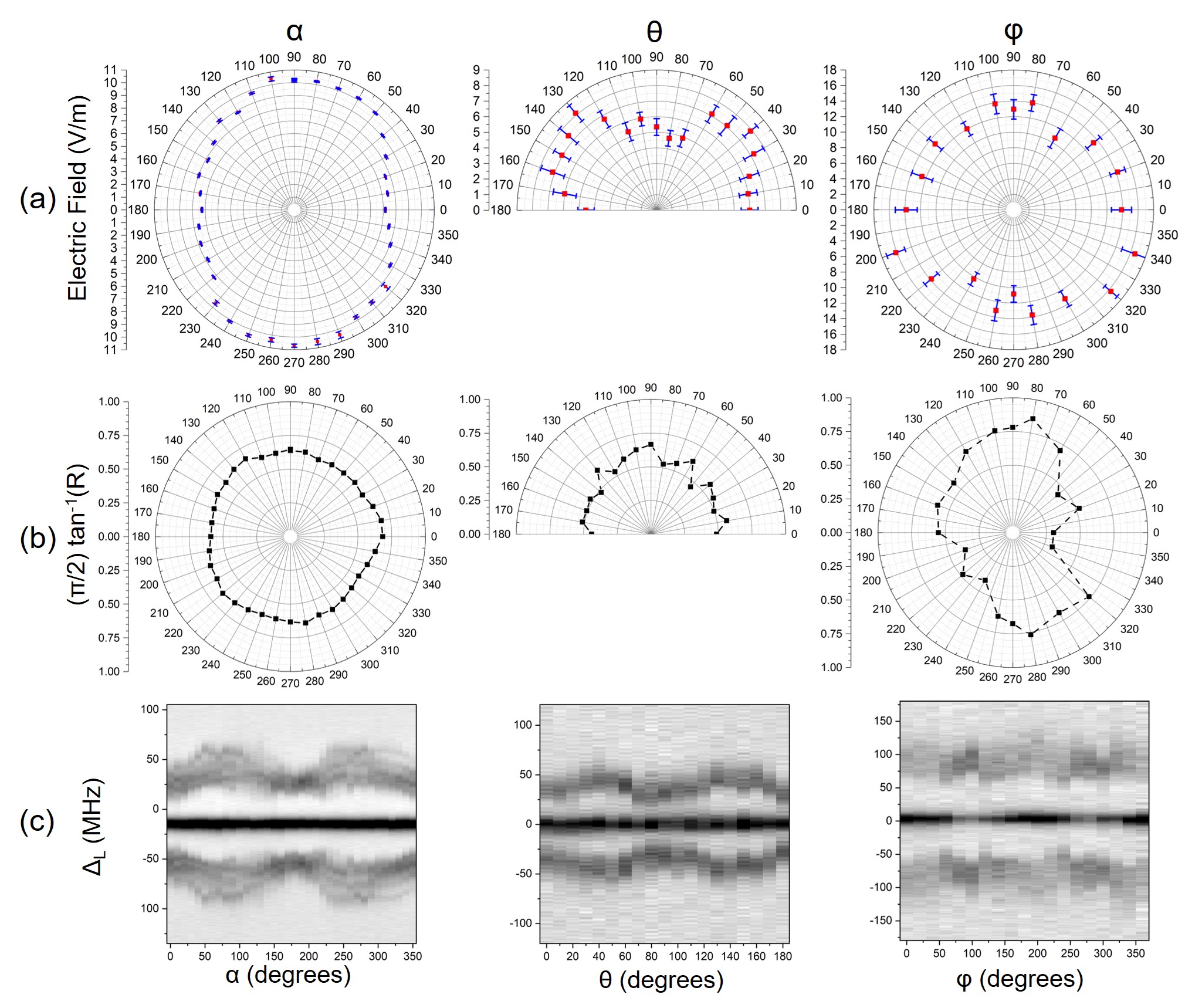}}
\caption{RFP field patterns measured for single-axis rotations  with rotation angles $\alpha$, $\theta$, and $\phi$ about the respective
primary axes $\hat{\bf{z}}$, $\hat{\bf{y}}$ and $\hat{\bf{x}}$ at 12.6~GHz RF. (a) RFP E-field patterns, (b) RFP polarization patterns using peak-height ratios $R$, (c) RFP atomic spectral output $S(\Delta \nu, *)$ from which E-field and polarization patterns are determined. In (b), we display the function $(\pi/2) \tan^{-1} (R)$ of the peak-ratio $R$.}
\label{fig3}
\end{figure*}

Figure~\ref{fig3} shows RFP field and polarization pattern measurements performed at 12.6~GHz RF for single-axis rotations of the RFP about the primary axes with rotation angles $\alpha$, $\theta$, and $\phi$ (see insets in Fig.~\ref{fig2}).  Figure~\ref{fig3}(c) shows the spectral outputs $S(\Delta \nu, *)$ of the RFP for $* = \alpha$, $\theta$, or $\phi$.
In the measurement, the range $180^o < \theta < 360^o$ is omitted from the characterization due to the presence of the RFP handle. The electric-field reception patterns of the RFP are obtained by implementing the $\langle E\rangle $ field determination method described in Sec.~\ref{sec3}-C. The corresponding RF-polarization patterns are expressed in terms of line-strength ratios, $R$, of the peaks in  $S(\Delta \nu, *)$.
The results are displayed in Fig.~\ref{fig3} rows (a) and (b), respectively.

In all RFP rotation planes we find field patterns with deviations from $4\pi$ isotropy due to mild RF-perturbations by RFP material structures surrounding the active atomic vapor. Detailed simulations of the RFP field perturbation, its effect on measurement uncertainties, and RFP self-calibration for SI-traceability using field pattern measurements and simulation results are presented in the subsequent sections.

The RFP polarization patterns are quantified by the ratio $R$ between the average area of the two AT-shifted peaks
(blue peaks in Fig.~\ref{figATAC} a) and the area of the central peak (white peak in Fig.~\ref{figATAC} a). The underlying physics is briefly explained in the following. The central peak in the AT spectrum in Fig.~\ref{figATAC} a corresponds to the magnetic sub-state $m_j=5/2$ of the Rydberg level 42$D_{5/2}$, where the direction of the RF-field's polarization defines the axis against which the $m_j$-value is measured in the quantum-mechanical solution of the problem. The $m_j=5/2$ level is not RF-shifted due to selection rules of the RF transition used in
Fig.~\ref{figATAC} a. The AT-shifted peaks (blue peaks in Fig.~\ref{figATAC} a) belong to the magnetic sub-levels $m_j=1/2,  \, 3/2$ components, with the AT shift of $m_j=1/2$ being $1.23$ times that of $m_j=3/2$. Note that in Fig.~\ref{figATAC} a the $m_j=1/2$ and $m_j=3/2$ components under the AT-shifted peaks are not resolved.
The line strengths of the cental ($m_j=5/2$) and AT-shifted ($m_j=1/2, \, 3/2$) components in the signal $S(\Delta \nu,*)$ are functions of the angle between the RF polarization and the polarization of the optical fields inside the RFP's vapor cell. Thus, the
line-strength ratio between the central peak and the AT-shifted peaks, displayed in row b in Fig.~6, is a measure for the RF polarization angle relative to the cell's (body-frame) $\hat{\bf{y}}$-axis (i.e., the direction of the optical probe fields). As a result, barring any imperfections in optical and RF polarizations, the line-strength ratio of central and AT-shifted peaks is insensitive in $\alpha$ and $\theta$, while it depends strongly on $\phi$. This behavior, born out in row b in Fig.~6, enables RF polarization measurements.

\begin{figure}[!t]
\centerline{\includegraphics[width=\columnwidth]{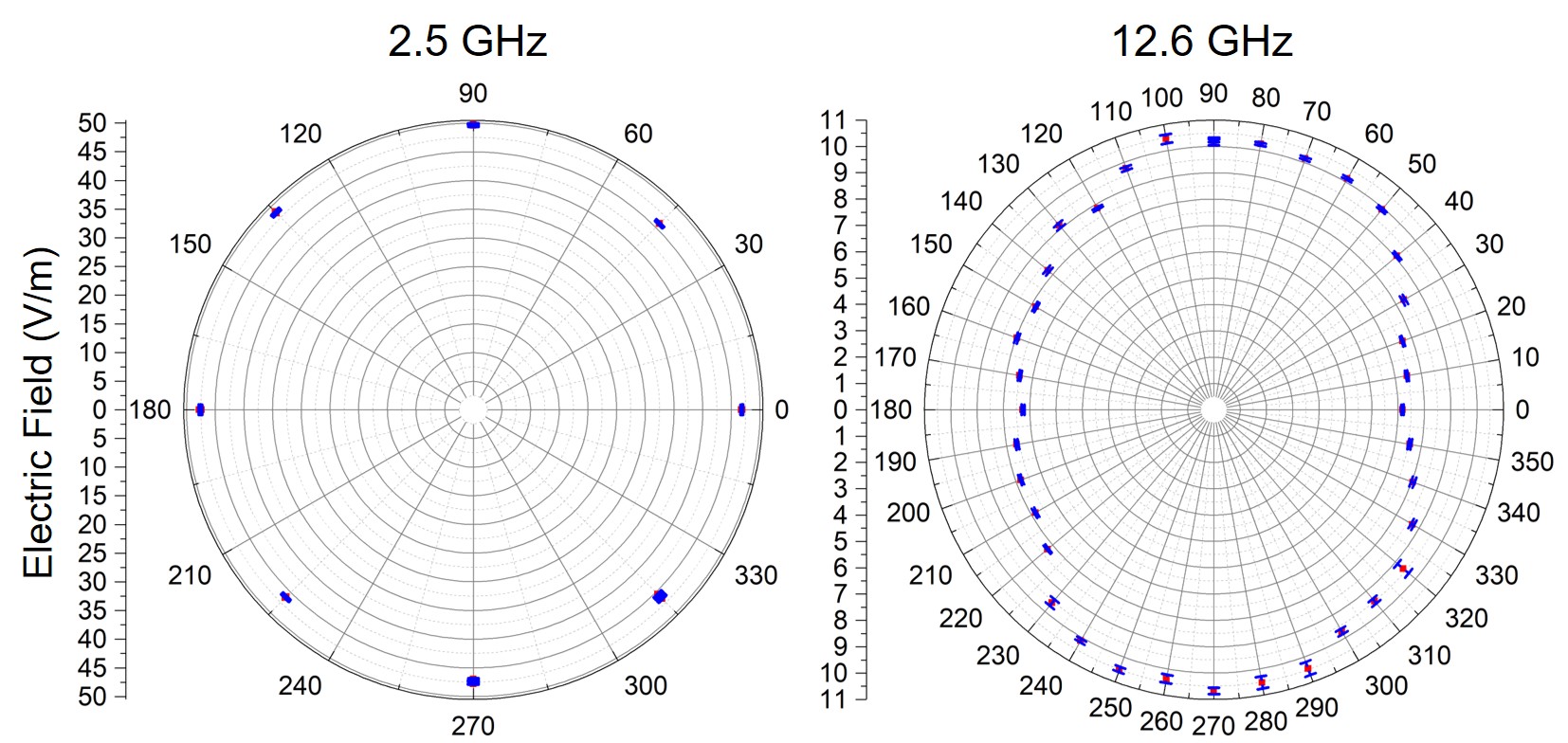}}
\caption{Comparison of RFP field patterns for single-axis rotation about the $\hat{\bf{y}}$-axis by angle $\alpha$ at 2.5~GHz (left) and 12.6~GHz (right).}
\label{fig4}
\end{figure}

For broadband atom-based RF measurements, the frequency dependence of the RFP response to the incident RF field must also be considered.  In Fig.~\ref{fig4} we show RFP field pattern measurements in $\alpha$ for 2.5~GHz RF, alongside the corresponding 12.6~GHz RF pattern. It is seen that the higher frequency has a stronger dependence on $\alpha$, as one might expect from the fact that shorter-wavelength RF fields are more prone to forming standing-wave patterns inside the RFP vapor cell. Regardless, even in the 12.6~GHz case the observed dependence is smooth and suitable for calibration.

\section{Finite-element simulations and RFP SI-traceable self-calibration}
\label{sec4p5}
One of the anticipated advantages of replacing antenna RF standards with Rydberg atom-based RF measurement is the ability to obtain a higher accuracy and reliability in the measurement by eliminating inevitable perturbations of the subject RF fields by the metal antenna probe used to measure the field.  The Rydberg atom-based RF sensing method alone provides a clear advantage in this regard, as the atoms represent perfectly frequency-matched quantum receivers for the incident RF, while there is negligible back-action of the atoms onto the incident RF field. However, the practical realization of an SI-traceable, self-calibrated Rydberg atom-based RF primary standard and Rydberg-based RF measurement instrumentation will require atomic RF probes necessarily comprising material structures to encapsulate the alkali vapor, guide and condition optical beams, and provide structural robustness and practical form factors for reliable use of the probe in testing environments.  Due to this, field perturbations by an atomic probe and inhomogeneous line broadenings of the atomic spectral signatures in the optical readout are unavoidable at some level over the ultra-wide band of RF frequencies accessible with the large variety of Rydberg-atom states that can be used.  As a result, to ensure accurate, traceable RF field measurements, RFPs must be pre-calibrated to account for perturbations of the RF E-field due to RFP geometry and material design choices, and the effect of these perturbations on the Rydberg atom spectral readout when performing atomic RF E-field measurements.

\begin{figure}[!t]
\centerline{\includegraphics[width=\columnwidth]{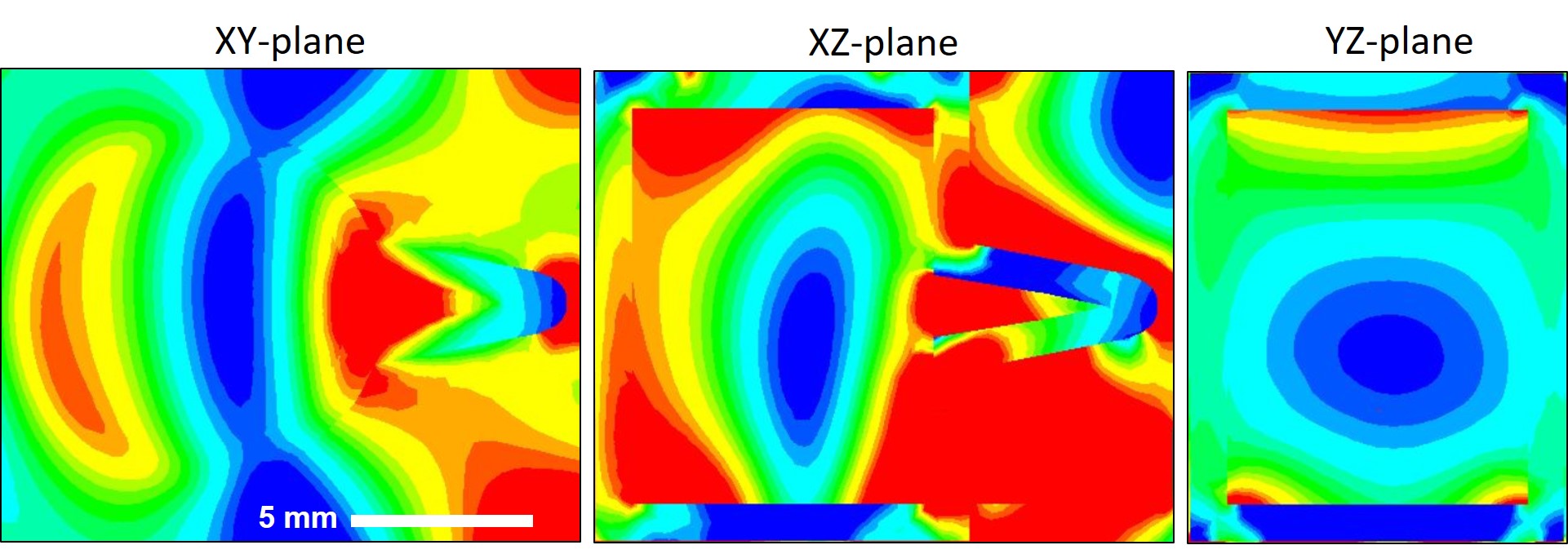}}
\caption{Simulated RF E-field internal to the RFP for a 12.6~GHz incident source as illustrated in Fig.~\ref{fig2}.  %Top row: Total RF E-field (left) and scattered field (right) in the $XY-$plane through the RFP.  Bottom row,
From left to right: Total RF E-field in the $YZ-$, $XZ-$, and $XY-$plane through the RFP. The field amplitude is displayed on a linear color scale ranging from 0.5 (blue) to 1.2 (red), in units of the incident plane-wave field amplitude. The ``ghost shapes'' visible in the images delineate the glass walls of the vapor cell used.    }
\label{fig5}
\end{figure}

The RFP presented here is designed with a geometry and low-dielectric-constant materials that afford both a small RF footprint and mechanical robustness for day-to-day use.  To characterize the effects of the dielectric material structures surrounding the optical detection region through the atomic vapor in the RFP, we perform finite-element simulations of the RFP for test points used in the field pattern measurements in Sec.~\ref{sec4}.  Simulation results for a 12.6~GHz RF plane wave incident on the RFP are shown in Fig.~\ref{fig5}.  The simulation considers a 12.6~GHz RF plane-wave source linearly-polarized along $\hat{\bf{z}}$ that is incident on the center of the cylindrical atomic vapor cell inside the probe ($\alpha$=0$\deg$).  The simulation accounts for all materials of the RFP including the rod and external housing (dielectric constant $\epsilon=2.6$), vapor-cell and embedded optics ($\epsilon=5.5$), and vacuum ($\epsilon=1$) defined inside the vapor-cell.  All RFP material component dimensions and positions in the simulation model are accurate to better than 1~mm.

\begin{figure}[!t]
\centerline{\includegraphics[width=\columnwidth]{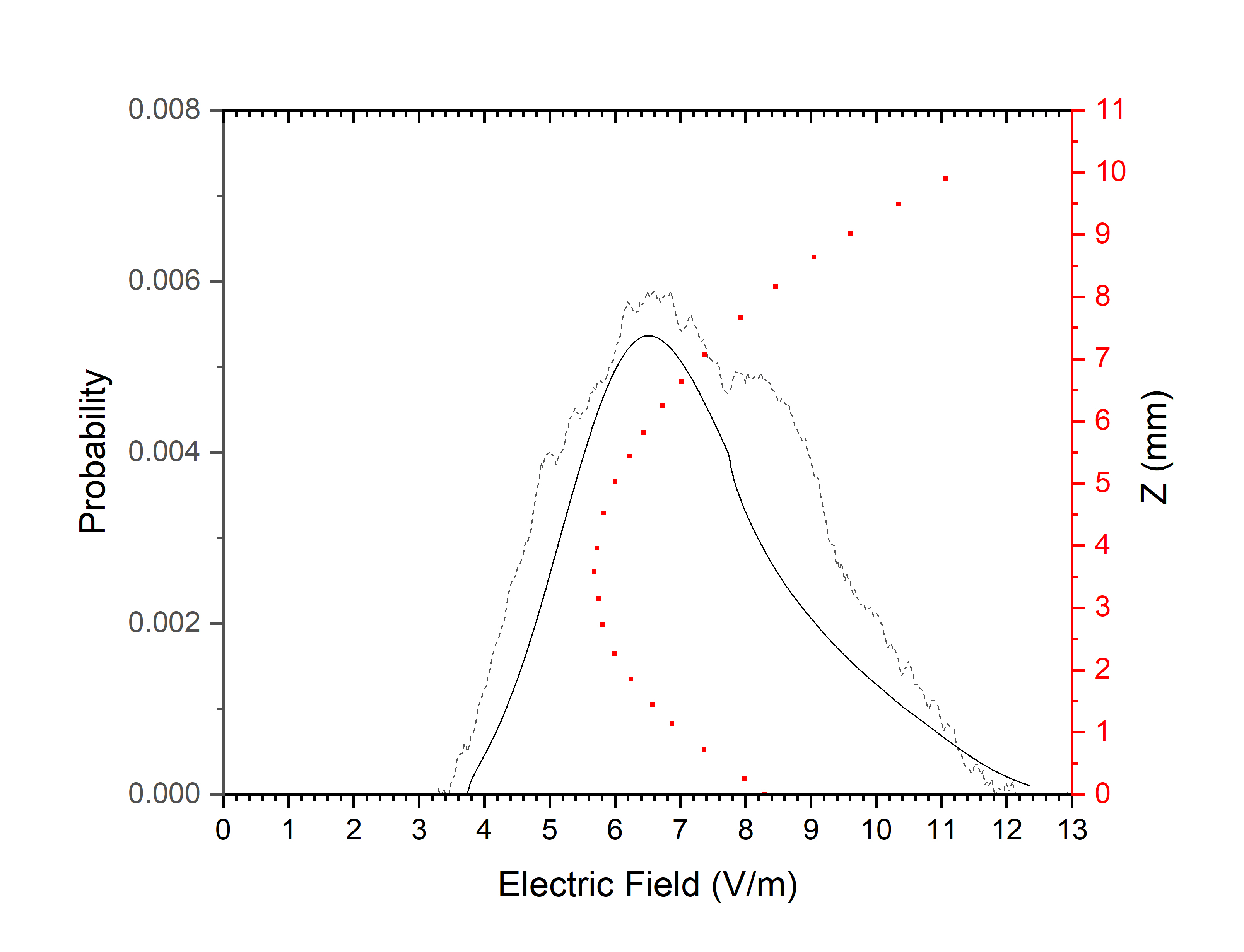}}
\caption{RF E-field probability distribution in the RFP atomic vapor along the optical beam path for a 12.6~GHz incident source.  Plotted are measured (black dashed) and simulated distributions (black solid), and the corresponding simulated spatial RF E-field distribution along $\hat{\bf{z}}$ (red dots).  }
\label{fig6}
\end{figure}

Figure~\ref{fig5} shows simulations of the incident 12.6~GHz RF field inside the RFP.  The field inhomogeneity in the optically-interrogated atomic-vapor detection region is due to the vapor-cell compartment being close to the size of the 12.6~GHz RF wavelength, while the asymmetry of the vapor cell geometry, such as its stem, has minimal effect.  To experimentally validate the simulation results, in Fig~\ref{fig6} we show simulated and measured 12.6~GHz RF field probability distributions along the optical beams propagating through the atoms in the $\hat{z}$-axis through the RFP vapor-cell.  A simulated spatial RF field distribution is also plotted.  The measured field distribution is extracted from the RF-modified EIT-AT lineshape from the RFP spectrum. The simulation reveals that the measured electric field has a distribution ranging from $\sim$0.55$\times$ to 1.1$\times$ the incident RF E-field.  The simulated and measured distributions are in very good agreement.

To quantify the effect of the enclosure-induced RF perturbation on the atom-based field measurement with the RFP we define a housing calibration factor $C = \langle E\rangle /E_{\rm{incident}}$, where $\langle E\rangle $ is the average internal field measured by the atoms and $E_{\rm{incident}}$ is the external, incident RF electric field.  From simulations for 12.6~GHz at $\alpha$=0$^{\circ}$ we obtain a C-factor of 0.71.  Using the C-factor and field patterns in Fig.~\ref{fig3} we fully characterize the RFP to provide self-calibrated E-field measurements directly SI-traceable to Planck's constant.  Following this methodology, RFPs for self-calibrated SI-traceable broadband operation implement angular- and frequency-dependent C-factor tables specific to the RFP model.  The initial, one-time characterization process implemented for the RFP model presented here for SI-traceability and self-calibrated operation in RF field measurement is generally applicable to other atom-based RF probes and detector types.

Due to the necessity of a compartment to hold the atomic vapor in RFPs, inhomogeneous line broadening of the atomic spectral signatures in the readout are unavoidable at some level over the ultra-wide band of RF frequencies accessible with Rydberg atoms.  In addition to engineering a given RFP probe to ensure minimal perturbations of the RF field over the desired RF field frequency and amplitude operating range, characterization and operation of RFPs require a means to account for changing RF field conditions and inhomogeneities in the detection volume during regular use of the instrument.  This is enabled by employing the spectral analysis method presented in Sec.~\ref{sec3} and employed in Sec.~\ref{sec4p5} for determination of the RF $\langle $E$\rangle $-field in the detection region.  RF measurement uncertainties associated with using this and other approaches in atom-based devices and probes such as the RFP is presented in the following section.

\section{Rydberg atom-based RF electric-field measurement uncertainty}
\label{sec5}
For application of the RFP instrument in RF metrology and the realization of self-calibrated, SI-traceable RF standard devices, an analysis and budget of measurement uncertainties when using a Rydberg-based measurement instrument is required.  A preliminary analysis of RF measurement uncertainties considering EIT linewidths and spectral features in the Autler-Townes regime has previously been discussed in~\cite{HollowayUncertainties2.2017} and characterizations of RF perturbations in the atomic detection volume due to the presence of dielectric vapor-cell materials and geometries have been investigated~\cite{Fan.2015,Anderson.2016,Zhang.2018}.  Further, an overview of fundamental factors contributing to RF sensitivity limitations using Rydberg EIT in atomic vapors for weak RF field sensing in the sub-Autler-Townes regime is presented in Ref.~\cite{Kuebler.2019}.  These provide valuable general insights into the limiting factors contributing to uncertainties in low RF E-field measurements with Rydberg EIT and Autler Townes splittings. However, they are insufficient for the realization of robust and practical Rydberg atom-based probes and instruments as SI-traceable RF standards and measurement tools suitable for use in real-world environments.  To this end, it is necessary to establish a general framework for a comprehensive uncertainty budget for atomic RF E-field measurements with real devices that accounts for uncertainty contributions from both the atomic measurement and the implemented analyses of atomic spectral signatures for the determination of the RF E-field, as well as systematics due to physical probe-device design attributes, such as the above-introduced C-factor, and back-end instrument hardware performance.

We present a comprehensive uncertainty budget and an overview of the factors contributing to uncertainties with an RFP instrument.  A detailed proposal and discussion of an uncertainty budget is provided in~\cite{Sapiro.2019}, intended to be generally applicable to atomic RF probes and devices employing Rydberg atom-based RF E-field measurement~\cite{Anderson2.2017} encompassing linear Autler-Townes splittings~\cite{Sedlacek.2012,Holloway.2014,Simons.2016}, AC Stark shifts~\cite{Anderson.2018}, and other non-linear regimes of the atom-RF interaction~\cite{Anderson.2014,Anderson.2016} for SI-traceable (self-calibrated) RF E-field measurement.  In Table~\ref{table_uncertainty} we present uncertainty budgets for several cases.
The uncertainty budgets are divided into two general classes of uncertainties: 1. Atomic-measurement uncertainties and 2. Probe-device uncertainties arising from external material and geometry design choices, as well as laser hardware stability during measurement.

The first two data columns in Table~\ref{table_uncertainty}  are for two different field determination analyses in the AT-splitting regime, as in the left panel of Fig.~\ref{figATAC}, the first based
on the expectation value of the field, $\langle E \rangle$, in the atomic detection region, and the second based on finding the dominant AT-shifted peak in the spectrum $S(\Delta \nu)$ and calculating the E-field for that peak, $E_P$.
The uncertainty analysis shows that the first method, which is discussed in some detail in Sec.~\ref{sec3} and is employed in the measurements performed in Section~\ref{sec4}, is more robust. The third column is for measurement of the RMS electric field, $E_{RMS} = \sqrt{\langle E^2 \rangle}$ using the quadratic AC Stark effect, as in the right panel of Fig.~\ref{figATAC}.

We note that there are operational uncertainties that ultimately contribute to any final E-field measurement, but that are under direct control of the operator and can, in principle, be eliminated. There is, for instance, a field-measurement uncertainty that results from the probe-to-source positioning uncertainty in the setup.  These types of systematic uncertainties affect E-field measurements performed with any class of probe device and are not intrinsic to the atomic-probe performance. Therefore, such operational uncertainties are not included in the uncertainty budget in Table~\ref{table_uncertainty}. They are, however, discussed and accounted for in the RFP field pattern measurements presented in Section~\ref{sec4}.

\begin{table}[!t]
\renewcommand{\arraystretch}{1.3}
% if using array.sty, it might be a good idea to tweak the value of
% \extrarowheight as needed to properly center the text within the cells
\caption{Uncertainty budget: Rydberg atom-based RF electric field measurements}
\label{table_uncertainty}
\centering
\begin{tabular}{|p{0.4\linewidth}||c|c|c|}
\hline
&AT $\langle E\rangle $ & AT $E_p$ & AC Stark $E_{\rm{RMS}}$ \\
\hline
Pixel & 0.1-0.3\% & 0.1-0.3\% & 0.3\%\\
\hline
Integration parameters & 0.7\% & NA & 0.3\%\\
\hline
Lineshape substructure & 0-6\% & 4-25\% & 1-3\%\\
\hline
\hline
Laser-frequency linearity & 0.2\% & 0.2\% & 0.2\%\\
\hline
\hline
Probe Calibration (C) factor  & 0.71 & 0.71 &  \\
\hline
\end{tabular}
\end{table}

\section{Pulsed detection and time-domain RF waveform imaging}

\label{sec6}
Many RF field detection and measurement applications require measurement of pulsed fields or modulated fields, in addition to continuous-wave (cw) fields that have been presented thus far in this paper. In order to meet these needs, the RFMS provides time-dependent field detection and RF wave-form imaging capability. In this mode of operation, the RFMS tracks the time-dependence of the atomic response received by the RFP. The time-dependent signal recordings, $S(\Delta \nu, t)$, reveal how the RF field depends on time, $t$.

\begin{figure}[!t]
\centerline{\includegraphics[width=\columnwidth]{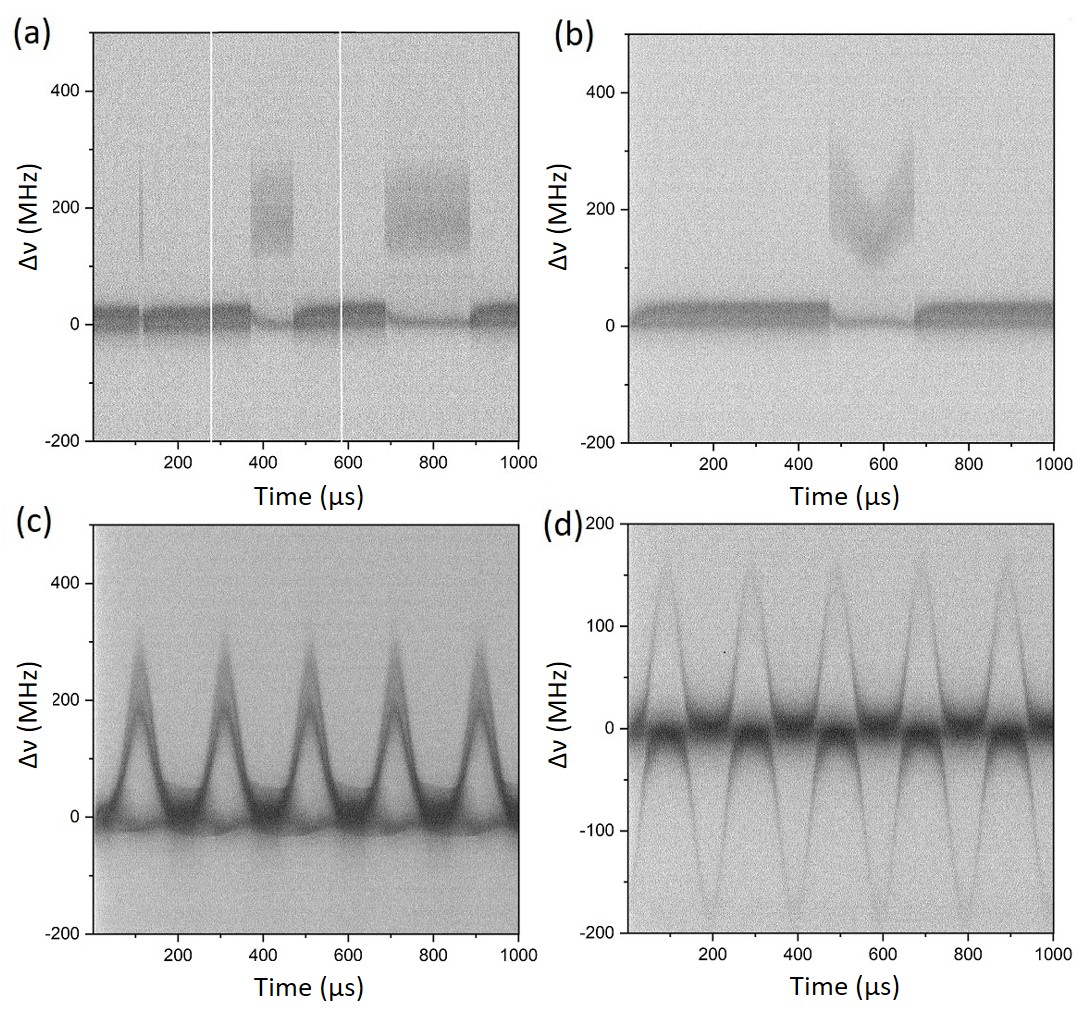}}
\caption{Records of time-dependent RF fields.  The recordings show RFP optical transmission displayed on a linear grayscale (arbitrary units), measured vs time and laser detuning. (a) EIT AC-Stark-shift response to square pulses of a 2.5-GHz RF signal with varying pulse lengths, from left to right, of 10~$\mu$s, 100~$\mu$s, and 200~$\mu$s with a repetition rate of 1~kHz. The RF intensity is constant during the pulses. (b) EIT AC-Stark-shift response to a pulsed signal that also is amplitude-modulated within the pulse. (c) Recording of a continuous, amplitude-modulated RF signal. The carrier frequency is 2.5~GHz and the AM baseband frequency is 5~kHz. (d) Recording of a continuous, frequency-modulated RF field near an AT resonance. The carrier frequency is 12.6 GHz, the FM baseband frequency is 5~kHz, and the FM deviation is 160~MHz.}
\label{fig8}
\end{figure}

Field detection of time-dependent fields using the RFMS and RF waveform imaging is demonstrated in Fig.~\ref{fig8}. The operating principle implements field determination methods similar to those used in the cw cases explained above, with spectral data recorded as a function of time. A 2.5~GHz carrier wave is incident on the RFP, with intensity varying in time. The time-dependent AC-Stark shift of the $42D_{5/2}$ state is recorded and displayed in real-time with a time resolution of 1~$\mu$s. In Fig.~\ref{fig8}~(a) we show recordings of square RF pulses with pulse lengths, from left to right, of 10~$\mu$s, 100~$\mu$s, and 200~$\mu$s; the pulse frequency is 1~kHz in all displayed cases. The square profile of the pulses is resolved to within the 1-$\mu$s time resolution used in the present demonstration. The RFMS can also resolve substructures within RF pulses; for example, in Fig.~\ref{fig8}~(b), we show a record of an RF square pulse with an overlaid sinusoidal amplitude modulation (10\% modulation depth, 5~kHz baseband frequency). In Fig.~\ref{fig8}~(c) we show a record of an amplitude-modulated cw RF signal (2.5~GHz carrier, 100\% modulation depth, 5~kHz baseband frequency). Figure~\ref{fig8}~(d) shows a record of an FM-modulated cw signal (12.6~GHz carrier, FM baseband frequency 5~kHz,  160~MHz peak deviation). The FM signal is probed using the $39D_{5/2}\rightarrow 40P_{3/2}$ AT resonance, which is resonant with the carrier.

\section{Conclusion}
\label{sec7}
In this work we have presented the first self-calibrating SI-traceable broadband Rydberg atom-based radio-frequency electric field probe (RFP) and measurement instrument (RFMS).  The RFMS is a commercial device comprising an atomic RF field probe (RFP), connected by a ruggedized fiber-optic patch cord to a portable mainframe control unit, with a computer software interface for probe RF measurement and analysis including real-time field and measurement uncertainty readout and spectral RF waveform visualisation. The RFP employs atom-based sensing using electromagnetically induced transparency (EIT) readout of spectral signatures from RF-sensitive Rydberg states of atoms in an atomic vapor~\cite{Anderson2.2017}.  The RFMS measures RF E-fields from resonant and off-resonant Rydberg-RF field interactions detected by the RFP probe head, and employs an RF-field-free atomic reference and an optical laser frequency tracker (OFT), which are integrated in the RFMS mainframe to ensure high reliability and precision in RF E-field measurements using Rydberg EIT spectroscopy in atomic vapors.  An overview of Rydberg EIT readout in atomic vapors for RF E-field measurement was provided and the operating principle of the RFP and RFMS was described.  An approach for the determination of the average RF E-field from spectral signatures in the RFP atomic-probe head has been introduced and implemented in demonstration RFP measurements of RF E-fields both near- and far-off-resonant from atomic Rydberg transitions. A complete characterization of an RFP probe is performed by measuring polar field patterns at 12.6~GHz RF, obtained by single-axis rotations of the RFP along primary axes in the far-field of a standard gain horn antenna.  Field pattern measurements at 2.5~GHz RF were also performed and RF polarization sensitivity demonstrated. We performed detailed finite-element simulations of the field inside the RFP at 12.5~GHz (Fig.~8) and 2.5~GHz (not shown) to quantify the effect of the probe's component materials and geometry on the RF E-field measurement by the optically-interrogated Rydberg atoms.  Simulation results were found to be in good agreement with the RFP field pattern measurements, revealing deviations of the RFP from a perfect isotropic RF receiver due to its specific materials and geometries.  The measurement and simulation results were in turn used to calibrate the probe referenced to the SI-traceable atomic RF E-field measurement. A Rydberg atom-based RF E-field measurement uncertainty budget and analyses were introduced and implemented in the RFP operation for SI-traceability of Rydberg atom-based RF probes and measurement tools in RF metrology.  Modulated and pulsed RF field measurement and detection capability with the RFP was also demonstrated and discussed. In extensive work not shown, broadband RF measurements of 3~MHz (HF-band) to $>$100~GHz sub-THz RF fields have also been performed.

The RFP instrument is a stand-alone device and new quantum technology platform with broad application potential.  In metrology, the RFP provides a first instrument suitable for use by metrology institutes worldwide for the establishment of a new atomic primary RF E-field standard by enabling the administration of round robin tests requiring standardized instrumentation and measurement methodology.  As a portable, broadband atomic RF E-field probe, the RFP is a single self-calibrated device that provides RF E-field measurement capability over an RF frequency range otherwise only accessible using multiple receiver antennas.  This can at once reduce the operational complexity, improve reliability, and reduce calibration costs in RF testing and measurement applications.  As a new platform technology, the RFP may be readily adapted to other application-specific RF sensing, receiving, and measurement needs and for the implementation of novel Rydberg-atom-based RF capabilities in communications, surveillance, and THz~\cite{Meyerpub.2018,Debpub.2018,RydbergRadio.2018,GetReadyForAtomicRadio,Song.2019,AndersonQuantumRadar.2019,Wade.2016}.

\section*{Acknowledgment}
This work was supported by Rydberg Technologies Inc.  Part of the presented material is based upon work supported by the Defense Advanced Research Projects Agency (DARPA) and the Army Contracting Command-Aberdeen Proving Grounds (ACC-APG) under Contract Number W911NF-17-C-0007.  The views, opinions and/or findings expressed are those of the author and should not be interpreted as representing the official views or policies of the Department of Defense or the U.S. Government.  The authors would like to thank Jeffrey McMahon for his assistance with the RF simulations presented in this work.

%\bibliographystyle{IEEEtran}
%\bibliography{IEEEabrv,RFMS_bibfile}

\end{document}